\documentclass[pra,amsmath,amssymb, twocolumn, showpacs]{revtex4}

\usepackage{dcolumn}
\usepackage[dvips]{graphicx}

\usepackage{epsfig}

\begin{document}

\title{Mapping two-qubit operators onto projective geometries}

\author{A.\ R.\  P. Rau$^{*}$}
\affiliation{Department of Physics and Astronomy, Louisiana State University,
Baton Rouge, Louisiana 70803-4001}


\begin{abstract}
The link between a quantum spin-1/2 and its associated su(2) algebra of Pauli spin matrices with Clifford algebra and quaternions is well known. A pair of spins, or qubits, which are important throughout the field of quantum information for describing logic gates and entangled states, has similarly an su(4) algebra. We develop connections between this algebra and its subalgebras with the projective plane of seven elements (also related to octonions) and other entities in projective geometry and design theory. 
   
\end{abstract}


\maketitle

\section{Introduction}

This paper establishes links between two entirely different areas of mathematics and physics. One is the manipulation of two spin-1/2, or four-level systems in quantum physics, which are of wide interest in quantum optics and in the field of quantum information \cite{ref1}. Apart from a unit $4 \times 4$ matrix, 15 linearly independent operators or matrices of the group SU(4), or algebra su(4), describe such systems. In the case of a single spin, or two-level, system, the similar set of three $2 \times 2$ Pauli matrices/operators of su(2) are very well-studied and completely familiar to all physicists, many also aware of their correspondence to quaternions \cite{ref2}, which are generalized numbers beyond reals and complex numbers. However, connection of su(4) to further generalized entities in mathematics such as octonions \cite{ref3} is little known. Octonions also have a connection to projective geometry, wherein points and lines obey a perfect duality, and to a whole branch of mathematics, ``Design Theory" \cite{ref4}, which has a fascinating history \cite{ref5}. 

We establish here connections between the operators of su(4) and these subjects. Specifically, each subalgebra of su(4) corresponds to a projective geometry and we give specific mappings of the operators of the subalgebra onto the projective diagram.  These subalgebras of su(4) occur in quantum logic gates and in quantum optics and molecular systems. Therefore, geometrical pictures for those su(4) operators involved may be useful while also suggestive generally for manipulations of a pair of entangled spin-1/2 (qubits), a key resource in quantum cryptography, quantum teleportation, and quantum computing \cite{ref1}. 

The arrangement of this paper is as follows. Section II gives a summary of quaternions and octonions, the latter naturally connecting to the projective plane of seven points and lines called the ``Fano Plane" \cite{ref3,ref4}. Section III links the algebra of su(4) and its various sub-algebras to diagrams in projective geometry. Thus the subalgebras su(2) $\times$ su(2) $\times$ u(1) and so(5), both important in quantum optics and quantum information, map onto projective geometries of seven and ten elements, respectively, while the full su(4) maps onto a fifteen-element projective diagram. The section ends by casting a very familiar problem of quantum physics, the hydrogen atom, in the same language.

\section{Quaternions and Octonions}
     
There are only four consistent arithmetics; more properly, ``real, normed division algebras" \cite{ref2}: reals, complex numbers, quaternions \cite{ref2}, and octonions \cite{ref3}. Associativity and commutativity of multiplication hold for reals and complex numbers, commutativity given up for quaternions, and even associativity lost for octonions although ``limited associativity" still holds, referred to as ``alternative" \cite{ref3}.

The first three of these have realizations in physics and are, indeed, ubiquitous. The elements of reality in classical physics, such as positions and momenta, are themselves observables, the results of measurement by our apparatus and our senses, and are perforce real. In quantum physics, the elements of reality are wave functions, which are complex numbers, not themselves accessible to measurement.  Bilinear combinations of them and their squared modulus give the observables, whether mean/expectation values or transition probabilities.

\begin{table}
\begin{center}
\begin{tabular}{|c||c||c||c||c||c||c||c|}

\hline
 $-$&$e_1$&$e_2$&$e_3$&$e_4$&$e_5$&$e_6$&$e_7$  \\ \hline
 \hline
$e_1$&$-1$&$e_4$&$e_7$&$-e_2$&$e_6$&$-e_5$&$-e_3$  \\
\hline
$e_2$&$-e_4$&$-1$&$e_5$&$e_1$&$-e_3$&$e_7$&$-e_6$  \\
\hline
$e_3$&$-e_7$&$-e_5$&$-1$&$e_6$&$e_2$&$-e_4$&$e_1$ \\
\hline
$e_4$&$e_2$&$-e_1$&$-e_6$&$-1$&$e_7$&$e_3$&$-e_5$ \\
\hline
$e_5$&$-e_6$&$e_3$&$-e_2$&$-e_7$&$-1$&$e_1$&$e_4$  \\
\hline
$e_6$&$e_5$&$-e_7$&$e_4$&$-e_3$&$-e_1$&$-1$&$e_2$  \\
\hline
$e_7$&$e_3$&$e_6$&$-e_1$&$e_5$&$-e_4$&$-e_2$&$-1$  \\
\hline
\end{tabular}
\end{center}
\caption{Multiplication table for octonions \cite{ref6}}
\end{table}

Extending to quantum spin, Pauli matrices $\sigma$ or Dirac spinors describe them. The algebra of Pauli matrices, $\sigma_i \sigma_j = \delta_{ij} + i \epsilon_{ijk} \sigma_k$, is in direct correspondence to that of quaternions. These three square roots of $-1$, denoted $(i, j, k)$, with $ij=k=-ji$ and cyclic, map into $-i\vec{\sigma}$. Attempts to formulate quantum mechanics with quaternions \cite{ref6} have not proven to have distinct advantages over the conventional formulation with complex numbers and Hermitian matrices. Already in classical physics, following Hamilton's invention of quaternions \cite{ref7}, Maxwell himself was inclined to use them for electromagnetism but vector calculus prevailed \cite{ref8}. Continued attempts to do classical mechanics with quaternionic concepts, termed ``geometrical mechanics" \cite{ref9}, have not had widespread adoption, although quaternions have many advantages since they constitute an algebra which vectors do not (division having meaning only among parallel vectors), making them particularly suited for describing rotations \cite{ref10}. In this regard, Gibbs has prevailed over Hamilton in physics \cite{ref3,ref8}!

Octonions (also called octaves or Cayley numbers \cite{ref3}), on the other hand, have remained esoteric with very limited attempts to apply them in physics \cite{ref11}, Baez putting it in colorful terms: ``quaternions are the eccentric cousin who is shunned at important family gatherings, octonions are the crazy old uncle nobody lets out of the attic" \cite{ref3}. They are defined by seven square roots of minus one, $(e_1, e_2, \ldots e_7)$, with a multiplication table as shown in Table I \cite{ref2,ref3}. 

Alternative arrangements exist in the literature but the above is fairly standard, has three minus and three plus signs in each row and column (not counting the diagonal), and if $e_i e_j =e_k$, this implies $e_{i+1} e_{j+1} =e_{k+1}$ and $e_{2i} e_{2j} =e_{2k}$. A more insightful, geometrical rendering is given by Fig.\ 1. Each of the triads on the seven lines (note the inscribed circle) obeys the rule that the product of two gives the third with a $\pm$ depending on whether it is in the direction of the arrow or opposed to it \cite{ref3,ref12}.

\begin{figure}
\vspace{-.5in}
\scalebox{1.6}{\includegraphics[width=2.in]{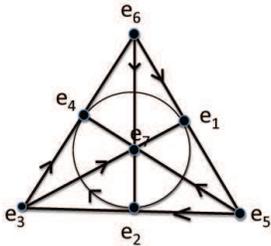}}
\vspace{-.4in}
\caption{The multiplication diagram for the seven octonions, also The Fano Plane of projective geometry with seven points and seven lines. The product of any two on a line equals the third with a $+/-$ depending on the direction of (along/against) the arrow \cite{ref2, ref3}.}
\end{figure}

This diagram of the multiplication among quaternions is also a central object of projective geometry, and is named after one of that subject's pioneers as ``The Fano Plane" \cite{ref3}. It has come to play an important role in coding theory and design theory \cite{ref4}. Each of the seven points lies on three and only three lines, and each line contains three and only three points. The occurrence of points and lines on an equal footing is an aspect of projective geometry. Unlike ordinary Euclidean geometry, it is characterized by such a ``duality" between points and lines, any valid theorem remaining true upon interchanging `point' and `line'. The Fano Plane is the smallest possible projective plane and is called a symmetric design or Steiner system in design theory \cite{ref4}. Alternative notations denote it as the projective geometry PG(2,2) or the Steiner triple or Balanced Incomplete Block (BIB) design with $v=b=7, k=3, \lambda =1$, written as 2-(7, 3, 1) \cite{ref4}.

As already noted, historical attempts to reformulate quantum physics with quaternions and octonions have not prevailed over the conventional formulation in terms of complex entities. It is not our intention in this paper to argue for any special virtue of these hypercomplex numbers in quantum physics. However, the 1:1 correspondence of various projective geometries with problems involving the operators of two quantum spins that we establish in the next section may well prove useful. At a minimum, our casting of sets of operators in the Clifford algebra of two qubits in the form of diagrams such as Fig.\ 1 is useful for expressing products and commutators of these operators and for keeping track of the results of successive operations on qubits in quantum information applications. These diagrams can, therefore, be seen as generalizations for two spins of the circle of $(i, j, k)$ that is familiar for Pauli spinors and quaternions.  

\section{Two spin-1/2 (qubits), the algebra and sub-algebras of su(4)}

A single quantum spin or qubit has the algebra of su(2) which is characterized by three parameters. The Pauli matrices $\vec{\sigma}$ provide a representation, and, as noted in Section II, $-i\vec{\sigma}$ maps 1:1 to quaternions. Today, in the field of quantum information (which embraces computation, cryptography, control and teleportation), a central object of interest is a pair of qubits \cite{ref1}. Logic gates for quantum computing and any entangled state of two sub-systems have such a pair as their basic element. It is ``the resource" of the field. The group SU(4) and its algebra su(4) describing such a pair of qubits has 15 generators/parameters. A convenient representation in terms of Pauli spinors $(\vec{\sigma}, \vec{\tau})$ and the unit matrix for each qubit is given by ($\vec{\sigma} \times \bf{I}^{(2)}$, $\bf{I}^{(1)} \times$ $\vec{\tau}$, $\vec{\sigma} \times$ $\vec{\tau}$). For explicit rendering in alternative forms and specific matrix representations, see \cite{ref13}-\cite{ref16}. Two prominent ones are as a direct product of two Pauli spinors \cite{ref13,ref14} or as the Gell-Mann basis for the su(4) algebra \cite{ref15,ref16}. 

Unlike the 1:1 correspondence between quaternions and $\vec{\sigma}$ of su(2), no similar rendering can be expected for the seven octonions and the fifteen su(4) matrices. Both the mismatch in the numbers and the lack of associative multiplication in the former while matrix multiplication is, of course, associative, preclude any exact correspondence. However, we will see close analogies between the various subalgebras of su(4) and the projective elements introduced in Section II as well as a looser connection between one subalgebra of seven operators and the octonion diagram of Fig. 1.

The algebra and subalgebras of su(4) have been throughly studied, with many applications in various areas of physics \cite{ref15,ref17}. A complete account of all the subalgebras is available in \cite{ref18}. In the context of the current intense investigations of quantum information, there are also many applications as in \cite{ref13,ref14,ref19}. For controllability of spin systems, successive Cartan decomposition and parametrization of higher su($2^n$) have also been studied \cite{ref20,ref21}. However, the connections we present here between these algebras and their operators with projective geometries do not seem to have been recognized except for recent, parallel, independent work by another group that came to our attention after our work was done \cite{ref22,ref23}. Interestingly, this group arrived at the connections from the other direction in terms of geometry and graph theory and not from the Lie algebras of multiple spins as we did. An illustration of the gains in overall understanding to be made by complementary approaches like this will be provided at the end of Sec. III C.      

\subsection{The subalgebra su(2) $\times$ su(2) $\times$ u(1) and The Fano Plane}
Among the subalgebras, an important one for applications in quantum information is su(2) $\times$ su(2) $\times$ u(1), the last term being a single element for a total of seven in this subalgebra. It describes \cite{ref13} the quantum controlled-NOT logic gate that been constructed with two Josephson junctions \cite{ref24}. The seven generators close under multiplication. That their commutators also close is exploited in solving for the time evolution of the system \cite{ref13}. There are several such distinct subsets of seven generators of su(4), one for each of the 15 generators \cite{ref13}. 

A specific example, with $\sigma_y \tau_y$ as the commuting u(1) element placed at the center, is shown in Fig.\ 2, the diagram arranged as an obvious analog of the Fano Plane in Fig.\ 1. That it is an analog, not an exact correspondence as already stated above, makes for some differences that we stress. Whereas the squares of the six outer elements is $-1$, that of the central one is $+1$. Also, arrows occur only on four of the lines, the three internal bisectors of the triangle having no directionality. Lines with arrows have the anti-commutativity aspect, that is, the product of two gives the third with an attached $\pm$ if in the direction of the arrow or against it. For the lines with no arrows, the three elements commute and the product of two gives the third, regardless of order. In physicists' language, one can say that there are four ``fermionic" and three ``bosonic" elements in the set of seven lines. The arrowed lines have the commutation relations of su(2) whereas the commutators of any two elements on the non-arrowed internal bisectors vanish.

\begin{figure}
\scalebox{1.5}{\includegraphics[width=2in]{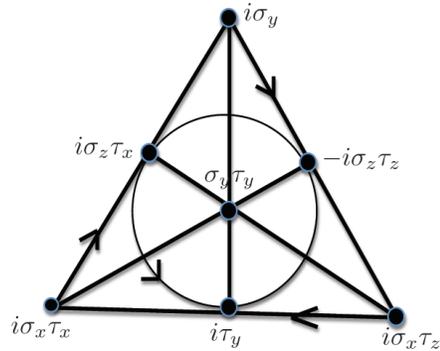}}
\vspace{-.2in}
\caption{One of sets of seven operators of two spin-1/2 that close under multiplication or commutators, arranged as in Fig.\ 1. They form the sub-algebra su(2) $\times$ su(2) $\times$ u(1) of su(4) \cite{ref13}.}
\end{figure}

A similar set of seven operators as in Fig.\ 2, except for interchanging $x$ and $y$, also constitutes a su(2) $\times$ su(2) $\times$ u(1) subalgebra and describes the Hamiltonian used in the experimental construction of the controlled-NOT gate with two Josephson junctions \cite{ref24}. We have analyzed this earlier in \cite{ref13} to provide density matrix elements as functions of the various parameters of the junctions. The multiplication table in that paper which was used extensively for the commutators of the seven operators involved in constructing the evolution operator can now be replaced by the more convenient Fig.\ 2.

\subsection{The subalgebra so(5) and Desargues's Theorem}
Another subalgebra of su(4) is so(5), the algebra of five-dimensional rotations, with ten generators. This too occurs in several physical situations in quantum optics and coherent population transfer in atomic and molecular systems involving four levels \cite{ref25}.  We analyzed such systems earlier in terms of an so(5) of ten operators which we now lay out in Fig.\ 3 as the ``Desargues" ten point/line figure of projective geometry, a striking, even marvellous, construct already as a figure \cite{ref3}. 

For any two triangles as in Fig.\ 3, arbitrarily oriented in space, with corresponding vertices connected by ``rays" from a point, the three points of intersection of corresponding sides lie on a common straight line. The two triangles bear a dual relationship to the point and to that lower line, and are said to be ``in perspective" from them \cite{ref3}. The connection to perspective in drawing and art is immediate and suggestive. (A variant, when the two (similar) triangles have parallel sides, will have those sides intersect at infinity, the points and line being at infinity \cite{ref26}, projective geometry making no distinction between parallel or intersecting lines.) Note that all the lines are arrowed. While not satisfying all the requirements of a projective plane as in Fig.\ 1, the Desargues construct is called a partial Steiner system in design theory.

As in the previous subsection, the merit of placing the operators of the Lie algebra in this projective geometrical figure lies in the patterns this suggests. Thus, consider the remark in the previous paragraph that moving the lowest line to become the line at infinity does not change its projective aspects. Interpreting this for the algebra so(5), this means removing dependence of the Hamiltonian on the three operators on the bottom line. Ten operators are still required for closing the so(5) algebra but the Hamiltonian depends only on seven parameters. Specifically, in our previous application to such so(5) systems in \cite{ref13}, this means setting the coefficients $F_{ij}$ equal to zero. How this affects the Hamiltonian and its associated time evolution and whether such systems may have value in quantum information is the kind of inquiry that the projective geometrical aspects point to. We hope to return to this elsewhere but it shows the possible usefulness of translating the projective geometry patterns into their corresponding realization in Lie algebra applications.   

\begin{figure}
\scalebox{1.5}{\includegraphics[width=2in]{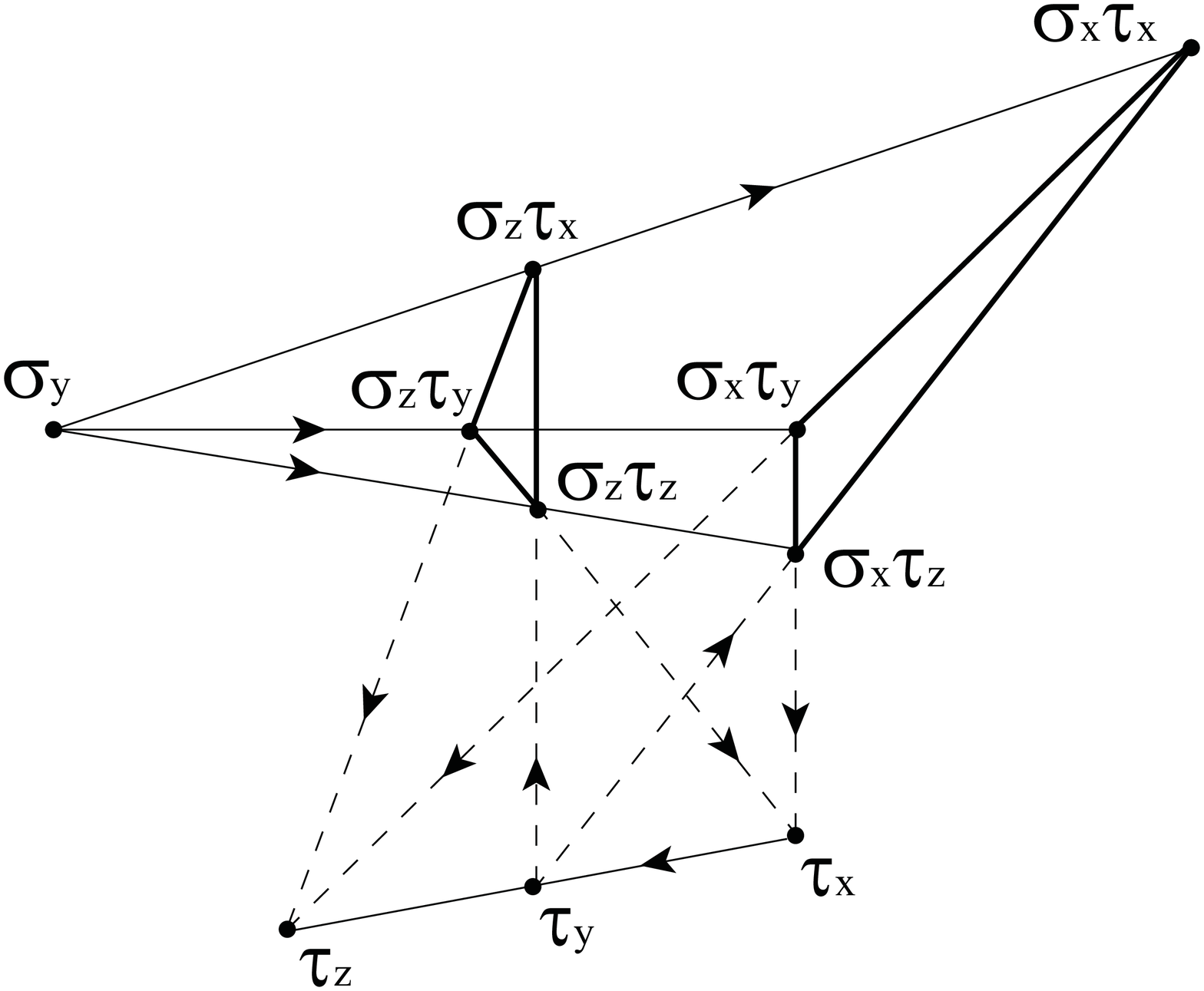}}
\caption{Ten operators of a two-spin system, forming the subalgebra so(5) of su(4), arranged as the Desargues diagram of projective geometry. The arrow notation is as in Fig.\ 1.}
\end{figure}

\subsection{The full su(4) algebra and its 15 point/line diagram}
Finally, and remarkably, all 15 generators of su(4) can be laid out using 15 straight lines as shown in Fig.\ 4. Only one of the lines has no directionality, corresponding to the three operators commuting among themselves. Again, while not a projective plane \cite{ref27}, this is a partial Steiner system of 15 elements. Just as in the variant of Fig. 3 with parallel triangles so that the three lowest points and the line they lie on recede to infinity, a more symmetrical variant of Fig. 4 can be drawn with seven pairs of parallel lines through 12 points. In that case, the top three points and the (unarrowed) line they lie on are at infinity.

\begin{figure}
\scalebox{1.5}{\includegraphics[width=2in]{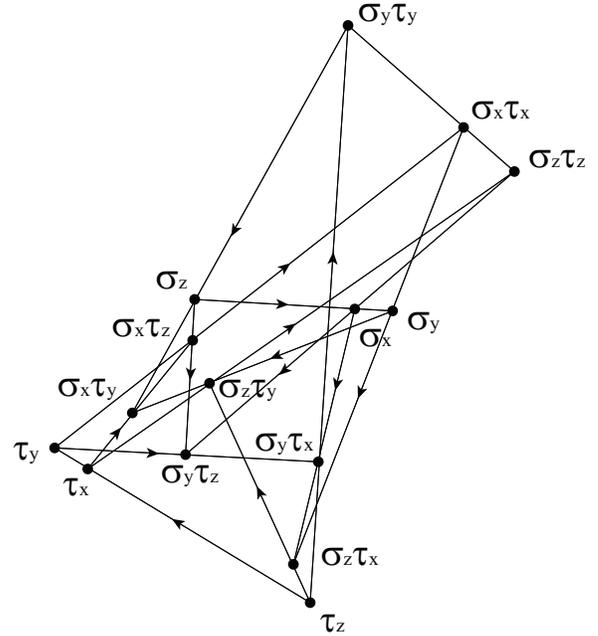}}
\vspace{-.1in}
\caption{The full set of fifteen operators of su(4) on a fifteen point, fifteen line diagram as in Figs.\ 2 and 3. One set of three mutually commute so that the top line carries no arrow sense.}
\end{figure}

An alternative rendering of Fig.\ 4 with a much more symmetrical look called the ``doily" is given in \cite{ref22,ref23,ref28}. As stated above, these authors have arrived at similar conclusions to ours but from a different starting point. Using graph-theoretical and geometrical studies of projective lines over the ring of 2 $\times$ 2 matrices and products of Galois fields GF(2), they have associated the 15 operators of su(4) with very symmetrical geometrical arrangements that are the counterpart of Fig.\ 4, called a Veldkamp space defined on a generalized quadrangle of order 2. They have also extended such geometrical studies to higher qudits. They consider a symplectic polar space $V(d, q)$ which is a $d$-dimensional vector space over a finite field $F_q$ with a bilinear alternating form. Denoted by $W_{d-1} (q)$, such a space exists only for $d=2N$, $N$ being called its rank. $W_{2N-1} (q)$ is the space of totally isotropic spaces of the projective geometry $PG(2N-1, q)$ with respect to a symplectic form. These authors identify the Pauli operators of $N$-qubits with the points of $W_{2N-1} (2)$.

As an illustration of different insights and simplifications possible with alternative points of approach, we consider one result in \cite{ref23}, namely, the number of points $v$ contained in $PG(2N-1, q)$ or $W_{2N-1} (2)$ which is $ v=2^{2N}-1 =4^N -1$. With the concept that two distinct points of $W_{2N-1} (2)$ are ``perpendicular" if they are joined by a line, $2^{2N-1}$ is the number of points that are not perpendicular to a given point. In terms of the Pauli operators of $N$-qubits, the property of ``commuting" translates to this concept of perpendicular. Together with a related number $D=v-1-2^{2N-1}$, this being the ``degree" of a graph with $v$ vertices, they provide a table of such numbers for various $N$ and remark that after posting their preprint, a physicist and a mathematician independently provided proofs of these results. 

However, from the perspective of the algebra of su($2^N$), these numbers are immediate and obvious. For $N=1$, any operator, say $\sigma_z$, does not commute with the other two Pauli matrices so that $v=3$ and $D=0$. For $N=2$, the case of two qubits and su(4), $\sigma_z$ commutes with all three $\vec{\tau}$ of the other spin and with the three bilinear ones $\sigma_z \vec{\tau}$ for a total of six so that $v=15, D=6$. This is the result noted and exploited in \cite{ref13} for the su(2) $\times$ su(2) $\times$ u(1) sub-algebra, that in tables of commutators such as in \cite{ref13,ref14}, there are six zeroes in each row besides the diagonal entry. For $N=3$ or three qubits, clearly the same enumeration extends: $\sigma^{1}_z$ commutes with all three $\vec{\sigma}^{2}$, all three $\vec{\sigma}^{3}$, the nine bilinear products of them $\vec{\sigma}^{2} \vec{\sigma}^{3}$, the six $\sigma^{1}_z \vec{\sigma}^{2}$ and $\sigma^{1}_z \vec{\sigma}^{3}$, and the nine $\sigma^{1}_z \vec{\sigma}^{2} \vec{\sigma}^{3}$ for a total $D=30$. This simple enumeration extends readily to give $D=126$ for $N=4$ and the result for general $N$, $D=v-1-2^{2N-1}$. 

Thus, counting from the algebraic commutation angle provides easily the number for the perpendiculars of projective geometry or the degree of a graph. This complements the last paragraphs in Sec. III A and III B where it was the projective geometrical picture which suggested simplifications in the counterpart Lie algebraic analysis. There are likely to be many such results which are more readily seen in one or the other approach of Lie algebras or projective geometries.    

\subsection{The hydrogen atom's SO(4) symmetry cast as The Fano Plane}
Another subalgebra of su(4) is su(2) $\times$ su(2) with six parameters/generators. Again, there are many such subsets of the fifteen su(4) operators, the most obvious being, of course, ($\vec{\sigma} \times \bf{I}^{(2)}$, $\bf{I}^{(1)} \times$ $\vec{ \tau}$). In an entirely different context than two spins, the well-known SO(4) symmetry of the hydrogen atom, with its six generators, the angular momentum $\vec{L}$ and the Laplace-Runge-Lenz vector $\vec{A}$, which close under commutation, $\vec{L} \times \vec{L} = i\vec{L}, \vec{L} \times \vec{A} =i\vec{A}, \vec{A} \times \vec{A} =i\vec{L}$, also affords another example \cite{ref29}. The linear combinations, $(\vec{L} \pm \vec{A})/2$, behave as two uncoupled angular momenta because $\vec{L} \cdot \vec{A} =0$, and provide a description of SO(4) as the product of two SO(3). These six operators, together with the unit operator, can therefore be arranged in Fig.\ 5 analogous to Fig.\ 2 and the Fano Plane in Fig.\ 1. Once again, the triad on the internal bisectors mutually commute and those lines carry no arrows.

\begin{figure}
\scalebox{1.5}{\includegraphics[width=2in]{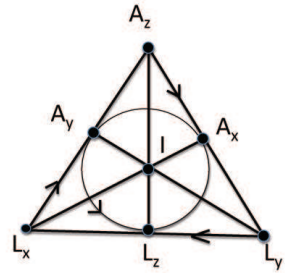}}
\vspace{-.4in}
\caption{The operators of angular momentum $\vec{L}$ and Laplace-Runge-Lenz vector $\vec{A}$, conserved quantities of the hydrogen atom, arranged along with the unit operator at the center as the Fano Plane of Fig.\ 1. }
\end{figure}

This rendering is of interest primarily as a curious analog, for the history and central importance of the problem, and for a historical ``connection" of human interest. The existence of another vector besides angular momentum in the Coulomb-Kepler problem goes back to Newton and Laplace, already in classical physics. Its quantum manifestation, and especially Pauli's initial solution of the hydrogen atom through the two commuting SO(3), is central to our understanding of the quantum-mechanical hydrogen atom. The su(2) $\times$ su(2) $\times$ u(1) construction in Fig.\ 2 is a close analog, except that it has a non-trivial commuting element at its centre instead of the unit operator in Fig.\ 5. And, finally the Fano Plane of these diagrams is named for the famous geometer G.\ Fano of the early twentieth century, whose son, U.\ Fano, became later in that century an eminent atomic physicist and who emphasized in his work and teachings symmetry principles including those in the SO(4) symmetry of the hydrogen atom \cite{ref30}. Fig.\ 5 now embeds that work in the diagram named for his father!

I thank Dr. Gernot Alber for discussions about The Fano Plane and for pointing me to \cite{ref4}, and Dr. Navin Singhi for discussions on projective spaces and design theory. I thank Dr. Alexander Rau for the figures and the Tata Institute of Fundamental Research, Mumbai, India, for its hospitality during the writing of this paper. And I am grateful to Dr. Metod Saniga for calling my attention to and sending copies of their work.


\begin{thebibliography}{}
\bibitem{ref12} See Baez in \cite{ref3} for a striking alternative that places the seven elements and unity at the corners of a cube. Planes through the origin (unity) give sub-algebras isomorphic to quaternions, lines through the origin give sub-algebras isomorphic to complex numbers and the origin itself the reals. Thus the Fano Plane is the projective plane over the two-element field $Z_2$ and is the smallest possible projective plane.
\end{thebibliography}
\end{document}